\documentstyle[12pt,aasms4]{article}

\begin{document}

\title{On the Nature of the X-ray Emission from M32}
\author{M. Loewenstein\altaffilmark{1}}
\affil{Laboratory for
High Energy Astrophysics, NASA/GSFC, Code 662, Greenbelt, MD 20771}
\altaffiltext{1}{Also with the University of Maryland Department of Astronomy}
\author{K. Hayashida and T. Toneri}
\affil{Department of Earth and Space Science, Osaka University,
Machikaneyama-cho, Toyonaka, Osaka 560, Japan}
\and
\author{D. S. Davis}
\affil{Center for Space Research, Massachusetts Institute of
Technology, Cambridge, MA 02139}

\begin{abstract}
We have obtained the first broad-band X-ray spectra of the nearby
compact elliptical galaxy M32 by using the {\it ASCA} satellite.
The extracted spectra and X-ray luminosity
are consistent with the properties of the hard spectral component 
measured in giant elliptical galaxies believed to
originate from X-ray binaries. Two {\it ASCA} observations were performed
two weeks apart; a 25\% flux decrease and
spectral softening occurred in the interval. We have also analyzed
archival {\it ROSAT} HRI data, and discovered that the X-ray emission
is dominated by a single unresolved source offset from the nucleus of M32.
We argue that
this offset, combined with
the extremely rapid large magnitude variations, and hard X-ray spectrum
combine to weakly favor a (single) X-ray binary over an AGN origin for the
X-rays from M32. The
nuclear black hole in M32 must be fuel-starved and/or
accreting from a radiatively inefficient advection-dominated disk:
the product of the accretion rate and
the radiative efficiency must
be less than $\sim 10^{-10}$ M$_{\odot}$ yr$^{-1}$ if
the X-ray source is indeed an
X-ray binary.
\end{abstract}


\section{Introduction}

At a distance of 700 kpc (\cite{t90}), 
M32 (NGC 221) is by far
the nearest elliptical galaxy to the Milky Way.
Although extremely compact, with a half-light radius of
$\sim 110$ pc and correspondingly low velocity dispersion
($\sigma\sim 80$ km s$^{-1}$) and optical luminosity
($M_B=-15.7$), M32 is structurally ``normal'' (\cite{k85})
and lies on the fundamental plane (\cite{b92}).
For these reasons it 
has played a foundational
role in studies of stellar populations and kinematics in elliptical
galaxies. In particular, M32 represents
the earliest, and one of the strongest, dynamical cases for the presence 
of a massive ($\sim 3\ 10^6 M_{\odot}$)
black hole in a non-active elliptical galaxy (\cite{b96}, \cite{v97}, and
references therein).
M32 is also a source of X-rays. More
than five magnitudes less optically luminous
than any other elliptical galaxy detected in X-rays, the
proximity of M32 presents a unique opportunity
to push the investigation of the nature of X-rays from 
elliptical galaxies to intrinsically very faint systems
and gain new insight into the ``hard component.''

It is now well-established (e.g., \cite{f89}) that
the X-ray emission from the brightest ellipticals is dominated
by hot gas that originated as stellar mass loss and subsequently
settled into hydrostatic equilibrium in the galactic potential.
However, there is a theoretical expectation (e.g., \cite{c91})
that less luminous galaxies have shallower gravitational potentials, so that
below some transitional luminosity the gas becomes unbound and
flows out in an unobservable galactic wind, thus revealing the
presence of the integrated emission from X-ray binaries. And indeed, the
hard X-ray flux from the ensemble of binaries -- circumstantial evidence
of which was discovered from {\it Einstein Observatory} 
observations (\cite{c87}; \cite{k92}) -- has been directly 
detected, first by BBXRT 
(\cite{s93}), and then
more universally by {\it ASCA} (\cite{a94}, \cite{m94}, \cite{m97}).

While, as expected, the fraction of the X-ray flux contained within
the hard component increases as the X-ray-to-optical flux ratio
decreases, the soft component proves to be surprisingly 
persistent and generally dominates the spectrum at energies less than 2
keV where {\it ASCA} and {\it ROSAT} are most sensitive.
In the 0.5-4.5 keV band, the 
luminosity of the hard component $L_{x,hard}\sim 4.1\ 10^{29}
(L_B/L_{B,\odot})$ erg s$^{-1}$, 
corresponding to a flux less than $5\ 10^{-13}$ erg cm$^{-2}$ s$^{-1}$
or only $\sim 70$ total (for four detectors, see next section) {\it ASCA}
source counts ks$^{-1}$ for a $10^{11} L_{B\odot}$ galaxy at the distance
of the Virgo Cluster (\cite{m97}).
Thus, the combination
of the large distance of even the closest giant ellipticals, the
intrinsic weakness of the hard component, and the obscuring effect
of the ubiquitous hot gas component severely limits what constraints
can be placed on the nature of the hard emission. Fits of thermal
models to {\it ASCA} spectra yield lower limits 
on the temperature of $\sim 2$ keV for individual galaxy spectra
and $\sim 6.5$ keV for a composite spectrum; however,
the data is equally well fit by a power-law with photon index $1.8\pm 0.4$
(\cite{m97}).
It is not clear whether and, if so, to what extent an active nucleus (AGN) is
contributing to the hard emission in some or all elliptical galaxies,
as might be expected if they host ``dead'' quasars with some
relic activity (\cite{f95}).
Because of the shallowness of its potential well, any gas bound to M32 will
have $kT<0.1$ keV and an unimpeded view of the hard component is
afforded. Despite this, previous X-ray studies have failed to
resolve the binaries/AGN ambiguity in the origin of X-rays from M32, 
with the latest attempt being a detailed
analysis of 
{\it ROSAT} PSPC observations by \cite{e96} (hereafter
EWD96). 

In this paper we present the results of analyses
of two recent {\it ASCA} observations of M32, as well as of
archival {\it ROSAT}
HRI data. 
Compared to the PSPC, {\it ASCA} has superior spectral energy resolution 
and sensitivity at energies greater than 2 keV, enabling us to derive more
accurate constraints and address the issue of spectral variability.
The excellent spatial resolution of the HRI provides a considerable
improvement in determination of the position and extent of the 
M32 X-ray emission. We have also examined archival 
{\it Einstein Observatory} and {\it EXOSAT}
data, and additional insight is sought from inspecting the long term
X-ray variability of M32. Although the evidence remains inconclusive,
we argue that the X-ray emission is likely dominated by a single
super-Eddington X-ray binary. If this is the case,
the very low upper limit on the luminosity associated
with accretion by the central black hole implies the presence of an
advection-dominated disk and/or fuel-starved AGN with an
extremely low value of the product of the accretion rate and
radiative efficiency.

\section{{\it ASCA} Observations and Data Analysis Procedures}

As described in more detail in \cite{t94},
{\it ASCA} has four identical, co-aligned X-ray telescopes and four focal
plane detectors: two gas imaging spectrometers (`GIS2' and `GIS3'), and
two solid state imaging spectrometers (`SIS0' and `SIS1'). Each SIS
consists of four CCD chips.
M32 was observed as part of the fourth round of the {\it ASCA} 
Guest Observer phase on 1996 July 25 (``observation A'')
and on 1996 August 4 (``observation B''), with the SIS in
single-CCD mode, and the GIS in PH mode. The GIS3 image from
observation A  
is shown in Figure 1, with the optical position of M32 
denoted by the `+' symbol. No other bright sources
in the field-of-view were anticipated based on previous observations
(see EWD96); however, surprisingly, two nearby sources -- each roughly
twice as bright as M32 -- are apparent. These objects evidently are
highly absorbed and/or highly variable hard sources, and are of
considerable interest in themselves. We defer discussion of their nature to
a separate paper (Toneri et al., in preparation).

Standard data screening procedures were followed (see \cite{d96}).
Data taken at elevation angles less than 10 degrees or
with cutoff rigidities below 6 GeV/c were excluded. For the
SIS, low bit-rate data and data taken at an elevation angle with
respect to the bright earth less than 20 degrees were also excluded.
Data were further cleaned using the usual rise-time discrimination
for the GIS and hot/flickering pixel removal for the SIS.

Spectra were extracted from approximately circular regions
in the plane of the sky mapped onto 
the detector plane in order to assure
accurate spectral response determination --
generated in the standard way for
point sources (\cite{d96}). Circles of radii $R\sim 3.0'$
and $R\sim 3.5'$ 
were chosen for SIS
and GIS observations, respectively, with areas lying
within $\sim 3.5'$ circular regions centered
on the other two bright sources excluded. 
The proximity of
the other point sources, and the chip boundaries for the SIS,
precludes the use of
larger extraction regions.
For each observation, 
we find accordance in 0.7-8 keV flux between detector pairs
at the $5\ 10^{-14}$ erg cm$^{-2}$ s$^{-1}$ level;  
joint spectral fits were not improved
by allowing relative normalizations to vary separately.
SIS (GIS) spectra were grouped to have a minimum
of 20 (25) counts per spectral bin, so that the $\chi^2$
statistic could be used  to determine best-fits and uncertainties.

For the SIS, there is insufficient proximate source-free area
for background calculation, and so standard blank sky
events lists obtained from the NASA/Goddard Space Flight Center
calibration database were used. For the GIS, we used
point-source-subtracted blank sky background spectra kindly
generated by K. Ebisawa. Because these files do not
take secular changes in the internal GIS background (\cite{mt})
into account, the background may be slightly underestimated in these files;
however, since best-fit spectral models
to GIS data are not flatter than the corresponding fits
to SIS data this 
evidently is a small effect. Use of local source-free regions
resulted in nearly identical spectral fits but with normalizations
lower by $\sim 20$\%.

The resulting exposure times and count rates used for spectral analysis
are shown in Table 1.

\section{{\it ASCA} Spectral Analysis Results}

As described above, we have extracted eight spectra that we will
subsequently refer to as G2A, G3A, S0A, S1A, G2B, G3B, S0B, and
S1B -- shorthand for GIS2 detector spectrum from observation A
(``G2A''), SIS1 detector spectrum from observation B
(``S1B''), etc.

The four spectra were fit individually, and
in various combinations,
for each observation to two distinct one-parameter spectral models:
a power-law continuum with photon index $\Gamma$
as a  model of AGN, and a thermal bremsstrahlung continuum
with temperature $kT$
as a model of single or ensembles of X-ray binaries. 
To optimize the statistical accuracy of the
results, simultaneous fits to all four detectors were carried out
for each observation.
In all cases, the best-fit parameters for individual detectors
are consistent, at better than 90\% confidence,
with the simultaneous fits. 

\subsection{Best-fit Parameters}

Spectrum S0A is shown in Figure 2,
along with the best power-law fit for observation A. There are
$\sim 960$ (background-subtracted) counts in this spectrum. In total,
there are $\sim 3200$ ($\sim 2300$) spectral
counts for observation A (B); therefore, the 
effective signal to-noise ratio is 1.8 (1.5) times that 
of the spectrum in Figure 2. Normalizations are tied together for
the GIS and SIS detector pairs for each observation (see above).
The best-fit parameters and 90\% confidence limits
are displayed in Table 2 for thermal and nonthermal fits to
each observation. Fits were performed 
with the column density fixed at
the Galactic value (6.3$\times$10$^{20}$ atoms cm$^{-2}$), as well as
with a freely varying
absorption column density. In the latter case, the best-fit columns are
consistent with the assumption of zero internal absorption and the
best-fit parameters only marginally change, although the uncertainties
increase slightly. The temperatures of the thermal model fits --
$\sim 9.5$-30 keV for observation A, $\sim 5.6$-13 keV for observation B --
and the photon indices of the nonthermal model fits -- 
$\sim 1.4$-1.6 for observation A, $\sim 1.6$-1.9 for observation B --
are consistent with the constraints on the composite spectrum
of the hard component in giant elliptical galaxies (\cite{m97}).
Unfortunately, because the M32 X-ray source was in a relatively
low-flux state during the {\it ASCA} observations (see below),
thermal bremsstrahlung and power-law models
provide equally good fits and are not statistically distinguishable.

\subsection{Variability Between {\it ASCA} Observations}

Because of the broad wings of the {\it ASCA} 
point-response function (PRF), there is a small but significant
contamination to the source counts at the position of M32 from
the relatively bright point source $\sim 5'$ away
(see Figure 1). The effect is more pronounced
for the GIS, since it produces additional
blurring to that of the {\it ASCA} X-ray Telescope alone (half-power radius
$\sim 1.5'$). 
Moreover, the level of this contamination varies slightly
from detector to detector
and between observations due to small differences in source
positions on the image plane relative to the optical axis of the telescope. 
Therefore, care must be taken in deriving the true M32 source flux, and
we have used the following procedure
to estimate 0.7-7 keV fluxes (averaged over the four detectors for
each observation). First, counting rates and spectra are 
extracted from a $3.5'$ radius circular region (in the detector plane)
centered on M32, and are used to derive a photon flux that consists of
contributions from both M32 and the nearest bright point source. Next, the
contamination component is estimated by multiplying the
photon flux of the second source by the ratio of effective areas
(averaged over the 0.7-7 keV band). These areas
have been convolved with the 
PRF at the position of each source and then integrated
over the M32 extraction region, and are computed in the same way
as the effective area curves used for spectral analysis. The
corrected flux is then the difference between the total flux 
and the contamination component ($\sim 20$\% of the total for the GIS,
$\sim 10$\%  for the SIS).
Our final estimated 0.7-7 keV fluxes for M32 are $1.35\ 10^{-12}$
erg cm$^{-2}$ s$^{-1}$ for observation A, and $1.05\ 10^{-12}$
erg cm$^{-2}$ s$^{-1}$ for observation B. The above procedure
was utilized for all four detectors, and the deviation from the
averages given above are less than $7\ 10^{-14}$
erg cm$^{-2}$ s$^{-1}$ for all detectors.
Comparison of the above fluxes with those corresponding to our
extracted spectra show that there is some residual contamination
in the latter. The effect of this is to artificially harden
the M32 spectra, since the nearby source is
slightly harder than M32 and the 
PRF is broader for higher energies. However, since the best fit models to the 
relatively uncontaminated SIS spectra are not 
softer than those
to the GIS spectra, this effect evidently is relatively unimportant.

In addition to the 25\% decrease in flux, there
is marginal evidence for spectral softening between observations A and
B. If the column density is assumed to be Galactic, the 
best-fit parameters differ at the 90\% confidence level; although,
there is marginal agreement if the columns are freely varied.
We
note that
spectral variability is only significant with the relatively small errors in
joint spectral fits to GIS and SIS data, and not in
either GIS or SIS fits independently.
If SIS spectra for both observations are fitted simultaneously
$\chi^2$ decreases by $\sim 7$-8 if $\Gamma$ or $kT$ are allowed to
vary separately compared to when they are constrained to be identical.
And, finally, there is a 
statistically significant ($>99$\%)
difference in hardness ratios ($R$), defined
as the 2-10 keV flux divided by the 0.5-2 keV flux -- $R=2.8\pm 0.13$
for observation A and $R=2.3\pm 0.14$
for observation B ($1\sigma$ errors). 

Using {\it ASCA} spectra, we derive fairly tight constraints
on the X-ray spectral parameters from a single observation.
EWD96 found it necessary to combine several {\it ROSAT} PSPC observations
in order to maximize the signal-to-noise ratio of their spectrum.
Reexamination of the {\it ROSAT} PSPC spectra 
reveals some 
evidence for spectral variability in these data as well --
the best fit power-law index (assuming the Galactic column density)
is 1.47($\pm 0.08$) for pointing ``WP600068'' (27 July 1991)
and 1.27($\pm 0.12$) for pointing ``WP600079'' (14 July 1991),
where the errors quoted are 90\% confidence uncertainties. 
Thus, the 
best-fits in EWD96 should be considered as luminosity-weighted
time-averages.

\section{{\it ROSAT} HRI Data Analysis}

M32 was observed with the {\it ROSAT} HRI for 12913 s spread out over
the week of 1994 July 19-26. Approximately 160 source counts were
detected (see below), which was sufficient to place significant constraints
on the position and extent of X-ray emission from M32.
Prior to spatial analysis, the data were flat-fielded and
corrected for particle background using the suite of programs
provided by the NASA/GSFC {\it ROSAT} Guest Observer Facility
(\cite{s95}). A contour map of the M32 emission,
superimposed on the optical image from the Digitized 
Palomar Observatory Sky Survey (DSS),
is shown in Figure 3;
the lowest contour corresponds to $\sim 10$ times the mean background level.

The position of M32 was estimated using both the sliding-cell method
in XIMAGE (\cite{a95}) and maximum-likelihood method
(``LDETECT'') in IRAF/PROS. The derived positions are 
consistent to better than $2''$. However, the image is slightly 
asymmetrically elongated (see below) and the image centroids and
peaks do not quite coincide.
We adopt a position $\alpha=00h42m 42.5\pm 0.2s$, $\delta=40d51' 52\pm 2''$.
The best-estimate position is consistent with the {\it Einstein}
HRI position (\cite{c84}) and is shown as an `X' in Figure 3.
The X-ray position is $\sim 10''$ offset from the optical position
published in the RC3 catalog  (\cite{d91}). The spatial coincidence of 
HRI and optical positions of a globular cluster in the halo of M31
supports the accuracy of this offset (see below).

Count rates are estimated using the XIMAGE ``SOSTA'' routine, and
``SRCINTEN'' within IRAF/PROS on the raw image. 
Both of these programs account for
effects due to
background, vignetting, and the PRF. 
We also utilized the 
IRAF/PROS ``IMCNTS'' program on the cleaned image. These methods yield
12.8, 13.6, and 13.3 cts ks$^{-1}$, respectively with a statistical
uncertainty of 1.2 cts ks$^{-1}$. The signal-to-background ratio
is $\sim 8:1$.
We adopt a count rate of $13\pm 2$ cts ks$^{-1}$, or $160\pm 25$ total
source counts (because of vignetting the effective
exposure time is not precisely equal to the actual duration
of observation).

XIMAGE produces a marginal ($\sim 3\sigma$) 
detection of a second point source
coincident with the optical position, and with a count rate
of $1.0\pm 0.36$ cts ks$^{-1}$. Moreover, the image is slightly elongated
in the east-west direction 
(not the direction of the spacecraft wobble)
and an azimuthally-averaged surface brightness
profile shows excess counts between $10''$ and $20''$ above what is expected
from the on-axis PRF. Figure 4 shows the results of subtracting 
a smoothed 
image corresponding to the re-normalized PRF from that shown in
Figure 3.
There is excess emission, significant at the $4\sigma$ level,
to the west of the overall emission peak.
Examination of the other point sources in the field using the same 
analysis method does not show any excess like that seen in M32. 
The peak of the M32 excess emission is nearly coincident with the optical
position, shown as a `+' symbol in Figure 4.  Although, 
these deviations from a point source centered on the optical position
are on the order of the HRI systematic aspect errors (\cite{m95}),  
they may indicate
that there is only a very weak source at the nucleus of M32,
with the bulk of the emission originating from one very luminous
X-ray binary near, but not at, the galactic center. The globular
cluster ``G144'' (source 15 in \cite{c84}) detected by the {\it Einstein} HRI
is also detected by the {\it ROSAT} HRI within $1''$ of the
optical position, lending confidence in the accuracy of our
derived X-ray position. (This position is also marked in the
GIS image -- Figure 1.)
Since the X-ray emission is distributed differently than the
stellar light (off-centered and more compact),
there seems to be no significant contribution from
a population of relatively faint unresolved discrete sources.

\section{Long and Short Term Variability}

Figure 5 shows the 17-year 0.2-2 keV light curve, including
{\it Einstein} IPC and HRI, {\it EXOSAT} CMA,
{\it ROSAT} PSPC and HRI, and 
{\it ASCA} observations. 
The PSPC data points are from EWD96 and correspond to distinct
observations of M31.
The conversions to the 0.2-2 keV band are made using the
{\it ASCA}-derived spectral constraints, and systematic uncertainties
associated with the emission model are included in the error bars. 
At the distance of M32, $10^{-13}$ erg cm$^{-2}$ s$^{-1}$
corresponds to $5.86\ 10^{36}$ erg s$^{-1}$. Since the 0.5-10 keV flux
is $\sim 3$ times the 0.2-2 keV flux, the luminosity is at times
well-above the Eddington luminosity for a 1 M$_{\odot}$ neutron star, with
a peak 0.5-10 keV luminosity of $\sim 6.7\ 10^{38}$ erg s$^{-1}$.
The flux varies over a full range of greater than 50, and can vary
by factors of several on timescales of days or less (EWD96).

We have also examined the M32 0.7-7 keV {\it ASCA} light curves, binned
into 1024 s intervals, to search
for variability on still shorter timescales. 
For all four detectors and for both observations,
the variations in
total (source-plus-background)
count rate are roughly
consistent with statistical fluctuations from a constant-flux 
source ($\chi^2$ per degree-of-freedom $=38/28$ and $18/27$ for
summed SIS and GIS 1024-bin lightcurves from observation A,
and $30/25$ and $30/21$ from observation B).
However, the (nearly constant) background level is significant and,
when subtracted for each bin, source-only light curves remain that are not 
formally consistent with a constant flux: variances are twice what
would statistically be expected. However, the variations are
often not coherent from detector to detector, and the statistical
significance of the variability is dependent on the details of how
source and background counts are extracted. 
As inclusion
of a modest systematic uncertainty would restore consistency with 
constancy, the evidence for short term variability currently is
marginal at best.
Background-subtracted coadded GIS and SIS 
light curves are shown in Figure 6.

No trend of hardness ratio with count rate is detected; 
however, the errors on the hardness ratio are large for the
low count rate intervals.

\section{Discussion}

While not definitively conclusive, the evidence points toward
an X-ray binary rather than an AGN origin for the X-ray source in M32.
The relatively short timescale
and large magnitudes of the
transient variations in flux (Figure 5)
are more characteristic of X-ray binary
than AGN behavior, and -- most significantly --
the X-ray
emission region seems to be centered $\sim 30$ pc from the center of M32.

The X-ray source in M32 is unlikely to be a high mass X-ray Binary
because of
the old stellar population of M32.
Among low mass X-ray Binaries (LMXBs), the X-ray temporal and spectral
characteristics most resemble high luminosity non-transient 
neutron star LMXBs
such as Sco X-1 or Cyg X-2 (\cite{l95}). Another possibility is
a binary containing a relatively massive ($>5$ M$_{\odot}$) black hole,
such as G2023+338. 

The extrapolation of the linear relationship between blue luminosity 
and the integrated X-ray luminosity from binaries leads to a predicted
0.5-4.5 keV luminosity of $\sim 10^{38}$ erg s$^{-1}$ for M32 --
comparable to the average observed luminosity. However, the hard
component is expected to be spatially distributed as the optical light
if it consists of many unresolved, faint, discrete sources. 
Such an extended component would easily be detected by the {\it ROSAT}
HRI since
it would produce $\sim 200$ counts inside the half-light radius
of $\sim 40''$ --
about nine times the background. Any unresolved component 
cannot be much more luminous
than $10^{37}$ erg s$^{-1}$.
That
the X-ray emission from M32 is more
centrally concentrated than the light
may simply be a consequence of the fact that,
in elliptical galaxies,
the integrated luminosity of discrete X-ray sources is dominated by
the brightest ($\sim 10^{38}$ erg s$^{-1}$) binary systems. If this is the
case, it would not be surprising if, at the low optical luminosity of M32,
a single binary dominates the X-ray emission and that it is located
near the galactic nucleus where the overall stellar density
is highly concentrated. The X-ray variability shown in Figure 5
implies that a single source dominates the emission, and not a compact
population of discrete sources. 
If the X-ray emission were instead associated with
an active nucleus (of an unusual kind), the ratio of integrated
X-ray binary luminosity to optical luminosity would be lower than what has
been measured
in other observed spheroidal systems (\cite{m97}).

If the X-ray emission in M32 indeed originates in an X-ray binary,
then the X-ray luminosity of any AGN is less than
$5\ 10^{36}$ erg s$^{-1}$ --
$\sim 10^{-8}$ of the Eddington luminosity of the central
massive black hole. 
(A conservative upper limit from the highest point in the
long-term light curve -- see Figure 5 -- is only a factor of $\sim 100$
greater.) 
This represents a uniquely low ratio
of X-ray luminosity to black hole mass 
($\sim 3\ 10^6 M_{\odot}$; \cite{v97}), compared with other objects
in the sample of \cite{h98}.
This is consistent with the lack of 
AGN indicators at other wavelengths (EWD96), and implies that
$\epsilon_x\dot M<10^{-10}$ M$_{\odot}$ yr$^{-1}$, where 
$\dot M$ is the accretion rate and $\epsilon_x$ the efficiency of
converting fuel into X-rays. 
Evidently, the black hole
is fuel-starved and/or has a highly radiatively
inefficient accretion disk. The integrated mass loss from
stars in M32 is
$\sim 0.005$ M$_{\odot}$ yr$^{-1}$. Since the central inflow
rate would be of this order for a global cooling flow, this 
would seem to argue against a lack of sufficient available fuel
for the central black hole. However, until there are
direct observational limits
on the central gas density in the relevant 50-100 eV
temperature range, this possibility cannot be excluded.

Radiatively inefficient,
advection-dominated accretion has been suggested as an explanation
for the relatively low X-ray luminosities in LINERs (e.g., 
NGC 4258, \cite{l96}), and for the paucity
of bright nuclear X-ray sources in elliptical galaxies 
that statistical arguments suggest should host
$10^8$-$10^9$ M$_{\odot}$ black holes (\cite{f95}, \cite{ma97}).
The ratio of X-ray luminosity to black hole mass in M32 is
greater than 1000 times
lower than in NGC 4258 if, as we have argued, the AGN is not the
primary X-ray source. The corresponding accretion rate for
an advection dominated disk with the `standard' structure
parameters of \cite{ma97}
is less than $2.5\ 10^{-4}(\eta_x/0.1)^{-1/2}$ in Eddington units, where
$\eta_x$ is the fraction of the disk luminosity emitted in the 
{\it ROSAT} band. Since this corresponds to $\dot M<1.6\ 10^{-5}
(\eta_x/0.1)^{-1/2}$ M$_{\odot}$ yr$^{-1}$, 
the M32 black hole cannot be fueled at the rate expected for
a global
cooling flow unless $\eta_x<10^{-6}$.
The low luminosity and
accretion rate in M32 is 
more in line with the advection-dominated
accretion disk model for the putative black hole in the center of our own 
galaxy (\cite{n95}) than with NGC 4258 or other LINERS.

\section{Summary}

We have attempted to illuminate the nature of the
X-ray emission from the nearby compact elliptical galaxy M32
by analyzing two newly obtained {\it ASCA} observations, as well as
archival {\it ROSAT} HRI data. 
The X-ray luminosity is consistent with 
the observed ratio between optical luminosity and integrated
emission from X-ray binaries observed in giant elliptical
galaxies; moreover,
the best-fit emission
models to {\it ASCA} spectra are consistent with the 
hard component detected in X-ray spectra of
giant ellipticals (characterized by $\Gamma\sim 1.8$ power-law or
$kT\sim 10$ keV bremsstrahlung models). Thermal and nonthermal models
are not distinguishable through analysis of the {\it ASCA} spectra.
The 0.7-7 keV luminosity
decreases from $\sim 8$ to $\sim 6\ 10^{37}$ erg s$^{-1}$, 
and the spectrum softens slightly
in the two weeks between {\it ASCA} observations.
The HRI image of M32 is dominated by unresolved emission 
apparently centered $\sim 10''$ (30 pc) to the east of the optical position,
although there is marginal evidence for excess emission at a position
consistent with the optical nucleus. There is no evidence
for diffuse, extended emission consisting of
unresolved discrete sources.
The X-ray/optical offset, combined with
the relatively rapid large magnitude variations, and hard spectrum
combine to weakly favor a single (Low Mass)
X-ray binary over an AGN origin for the
X-rays from M32. More precise identification of the source
({\it i.e.} as either a neutron star or black hole LMXB) requires
higher quality spectral and temporal data.
If the X-ray source is an
X-ray binary, then
the nuclear black hole in M32 must be fuel-starved and/or
accreting from a radiatively inefficient advection-dominated disk with
$\epsilon_x\dot M<10^{-10}$ M$_{\odot}$ yr$^{-1}$, and the
M32 nucleus has the lowest
X-ray luminosity-to-mass ratio of any identified black hole.

\acknowledgments
 
We have made
extensive use of the HEASARC data base, and SKYVIEW, XANADU
(XSPEC, XRONOS, XIMAGE), FTOOLS,
IRAF/PROS, and IDL software packages. ML wishes to thank
L. Angelini, K. Ebisawa, K. Gendreau, U. Hwang, R. Mushotzky, A. Ptak, and
T. Yaqoob for advice and assistance. This paper benefited 
from a constructive referee's report.

\clearpage

\begin{deluxetable}{cccc}
\footnotesize
\tablecaption{{\it} ASCA Spectra of M32} 
\tablewidth{530pt}
\tablehead{
\colhead {Observation} &
\colhead {Detector} &
\colhead {Good time} &
\colhead {Count Rate\tablenotemark{a}}\\
\colhead {} &
\colhead {} &
\colhead {(s)} &
\colhead {(cts ks$^{-1}$)} }
\startdata

A & GIS2 & 28848 & 22.7 \nl
A & GIS3 & 28850 & 26.7 \nl
A & SIS0 & 28086 & 34.2 \nl
A & SIS1 & 28116 & 29.0 \nl
\hline
B & GIS2 & 28372 & 17.7 \nl
B & GIS3 & 28368 & 23.3 \nl
B & SIS0 & 22963 & 26.8 \nl
B & SIS1 & 22791 & 23.3 \nl

\enddata

\tablenotetext{a} {Spectra were extracted over
the 0.7-8 (0.6-7.5) keV energy band for GIS (SIS) spectra;
count rates are background subtracted.}
 
\end{deluxetable}

\clearpage

\begin{deluxetable}{cccc}
\footnotesize
\tablecaption{Best-fit Parameters and 90\% Confidence Errors from
Joint SIS+GIS Spectral Analysis}
\tablewidth{530pt}
\tablehead{
\colhead {obs.} &
\colhead {$N_H$\tablenotemark{a}} &
\colhead {$\Gamma$ or $kT$\tablenotemark{b}} &
\colhead {$\chi^2/\nu$\tablenotemark{c}} }
\startdata

A & 6.3* & $1.49_{-0.06}^{+0.07}$ & 154/154 \nl
A & $7.6_{-7.5}^{+8.9}$ & $1.51_{-0.14}^{+0.15}$ & 154/153 \nl
B & 6.3* & $1.68_{-0.08}^{+0.08}$ & 124/122 \nl
B & $10_{-8.7}^{+10}$ & $1.74_{-0.18}^{+0.20}$ & 123/121 \nl
\hline
A & 6.3* & $12.6_{-2.8}^{+5.1}$ & 153/154 \nl
A & $2.9_{-2.9}^{+6.8}$ & $15.4_{-5.9}^{+14}$ & 152/153 \nl
B & 6.3* & $6.94_{-1.29}^{+1.90}$ & 128/122 \nl
B & $1.4_{-1.4}^{+7.9}$ & $8.23_{-2.62}^{+4.57}$ & 127/121 \nl

\enddata

\tablenotetext{a} {Column density in units of $10^{20}$ cm$^{-2}$;
asterisk denotes column density fixed during spectral fitting.}
\tablenotetext{b} {Best-fit photon index for power-law model
fits (initial four entries), or temperature for
thermal bremsstrahlung model fits (final four entries); also
shown are $\chi^2+2.7$ uncertainties for one-parameter fits,
$\chi^2+4.6$ uncertainties for two-parameter fits.}
\tablenotetext{c}{ Minimum value of $\chi^2$ per degree of freedom.}
 
\end{deluxetable}

\clearpage

\figcaption{GIS3A image of M32 field, smoothed with a $30''$ Gaussian.
M32 is the faintest, southernmost of the three point sources. The cross
denotes the optical position of M32, the star that of a globular
cluster in the halo of M31.
The contour levels are 1.1, 2.2, 3.3, 4.4, 5.5, and 6.6
counts pixel$^{-1}$ (1 pixel $=15''\times 15''$)  -- 
the lowest contour is at about 2.5 times the average background
level. 
1 count pixel$^{-1}$ corresponds to $\approx 4.7\ 10^{-14}$ erg 
cm$^{-2}$ s$^{-1}$ arcmin$^{-2}$ in the {\it ASCA} (0.5-10 keV) bandpass.}

\figcaption{S0A spectrum (rebinned for presentation)
with power-law model that provides the best simultaneous fit
to G2A, G3A, S0A, and S1A spectra (histogram).}

\figcaption{Linearly spaced {\it ROSAT} HRI contours (in white)
superimposed on the DSS optical image of M32. 
The HRI image has been flat-fielded, background-subtracted, and
smoothed with a $5''$ Gaussian. The contour levels are at 1, 4, 7 and 10
counts pixel$^{-1}$ (1 pixel $=5''\times 5''$); the average background
level is $\approx 0.1$ counts pixel$^{-1}$; and,
1 count pixel$^{-1}$ corresponds to $\approx 3.8\ 10^{-13}$ erg 
cm$^{-2}$ s$^{-1}$ arcmin$^{-2}$ in the {\it ROSAT} (0.2-2 keV) bandpass.
The `+' denotes the
optical and the `X' the X-ray position of M32. Since the HRI contours
deviate from point-like emission, the X-ray centroid and peak do not
precisely coincide.
The GIS3a
contours from Figure 1 have been replotted (in black).}

\figcaption{Same as Figure 3, but now with the smoothed image
of the renormalized on-axis HRI PRF subtracted. The contour levels are
now 1 and 2 counts pixel$^{-1}$. The significance of this feature
is $\sim 4\sigma$.}

\figcaption{M32 Lightcurve -- flux in the PSPC band 
(0.2-2 keV) vs. time --
encompassing {\it Einstein}, {\it EXOSAT}, {\it ROSAT} and {\it ASCA}
observations. The solid lines denote the fluxes corresponding to the 
1$M_{\odot}$ Eddington luminosity (``Ledd'') and one-third Ledd at the
distance of M32 -- the bolometric X-ray luminosity is at least three
times the 0.2-2 keV value.}

\figcaption{Background-subtracted, co-added {\it ASCA} GIS and
SIS M32 lightcurves for observations A and B.}


\clearpage

\begin{thebibliography}{}
\bibitem[Angelini et al. 1995]{a95}
Angelini, L. 1995, XIMAGE Users's Guide for version 2.53\\
($http://heasarc.gsfc.nasa.gov/docs/xanadu/ximage/ximage.html$)
\bibitem[Awaki et al. 1994]{a94}
Awaki, H. et al. 1994, \pasj, 46, L65 
\bibitem[Bender et al. 1992]{b92}
Bender, R., Burstein, D., \& Faber, S. M. 1992, \apj, 399, 462
\bibitem[Bender et al. 1996]{b96}
Bender, R., Kormendy, J., \& Dehnen, W. 1996, \apj, 464, L123
\bibitem[Canizares, Fabbiano, \& Trinchieri 1987]{c87}
Canizares, C. R., Fabbiano, G., \& Trinchieri, G. 1987, \apj, 312, 503
\bibitem[Ciotti et al. 1991]{c91}
Ciotti, L., D'Ercole, A., Pellegrini, S., \& Renzini, A. 1991, \apj, 
376, 380
\bibitem[Crampton et al. 1984]{c84}
Crampton, D., Cowley, A. P., Hutchings, J. B., Shade, D. J., \&
Van Speybroeck, L. 1984, \apj, 284, 663
\bibitem[Day et al. 1996]{d96}
Day, C. S. R., Arnaud, K. A. A., Ebisawa, K., Gotthelf, E. V.,
Ingham, J., Mukai, K. \& White, N. E. 1996, The ABC Guide to
{\it ASCA} Analysis 
($http://heasarc.gsfc.nasa.gov/docs/asca/abc\_contents.html$)
\bibitem[de Vaucouleurs et al. 1991]{d91}
de Vaucouleurs, G., de Vaucouleurs, A., Corwin, Jr., H. G., Buta, R. J., 
Paturel, G., \& Fouqu\'e, P. 1991, 
Third Reference Catalogue of Bright Galaxies (New York: Springer-Verlag)
\bibitem[Eskridge, White, \& Davis 1996]{e96}
Eskridge, P. B., White, R. E., III, \& Davis, D. S. 1996, \apj, 463, L59
(EWD96)
\bibitem[Fabbiano 1989]{f89}
Fabbiano, G. 1989, \araa, 27, 87
\bibitem[Fabian \& Rees 1995]{f95}
Fabian, A. C., \& Rees, M. J. 1995, \mnras, 277, L55
\bibitem[Hayashida et al. 1998]{h98}
Hayashida, K., Miyamoto, S., Kitamoto, S., Negoro, H., \& Inoue, H. 1998,
ApJ, in press
\bibitem[Kim, Fabbiano, \& Trinchieri 1992]{k92}
Kim, D. -W., Fabbiano, G., \& Trinchieri, G. 1992, \apj, 393, 134
\bibitem[Kormendy 1995]{k85}
Kormendy, J. 1985, ApJ, 295, 73
\bibitem[Lasota et al. 1996]{l96}
Lasota, J.-P., Abramowicz, M. A., Chen, X., Krolik, J., Narayan, R., \&
Yi, I. 1996, \apj, 462, 142
\bibitem[Lewin et al. 1995]{l95}
Lewin, W. H. G., Van Paradis, J., van den Heuvel, E. P. J. 1995,
X-ray Binaries (Cambridge: Cambridge University Press)
\bibitem[Mahadevan 1997]{ma97}
Mahadevan, R. 1997, \apj, 477, 585
\bibitem[Matsumoto et al. 1997]{m97}
Matsumoto, H., Koyama, K., Awaki, H., Tsuru, T., Loewenstein, M., \&
Matsushita, K. 1997, \apj, 482, 133
\bibitem[Matsushita et al. 1994]{m94}
Matsushita, K. et al. 1994, \apj, 436, L41
\bibitem[Miyata 1997]{mt}
Miyata, E. 1997, Ph. D. thesis, University of Osaka
\bibitem[Morse et al. 1995]{m95}
Morse, J. A., Wilson, A. S., Elvis, M., \& Weaver, K. A. 1995, 
\apj, 439, 121
\bibitem[Narayan, Yi, \& Mahadevan 1995]{n95}
Narayan, R., Yi, I., \& Mahadevan, R. 1995, \nat, 374, 623
\bibitem[Serlemitsos et al. 1993]{s93}
Serlemitsos, P. J., Loewenstein, M., Mushotzky, R. F., Marshall, F. E.,
\& Petre R. 1993, \apj, 413, 518
\bibitem[Snowden 1995]{s95}
Snowden, S. L. 1995, Cookbook for Analysis Procedures for {\it ROSAT}
XRT/PSPC Observations of Extended Objects and the Diffuse Background\\
($ftp://legacy.gsfc.nasa.gov/rosat/software/fortran/sxrb/cookbook.tex$)
\bibitem[Tanaka, Inoue, \& Holt 1994]{t94}
Tanaka, Y., Inoue, H., \& Holt, S. S. 1994, PASJ, 46, L37
\bibitem[Tonry, Ajhar, \& Luppino 1990]{t90}
Tonry, J. L., Ajhar, E. A. \& Luppino, G. A. 1990, \aj, 100, 1416
\bibitem[van der Marel et al. 1997]{v97}
van der Marel, R. P., de Zeeuw, P. T., Rix, \& H. W., Quinaln, G. D. 1997,
Nature, 385, 610
\end{thebibliography}
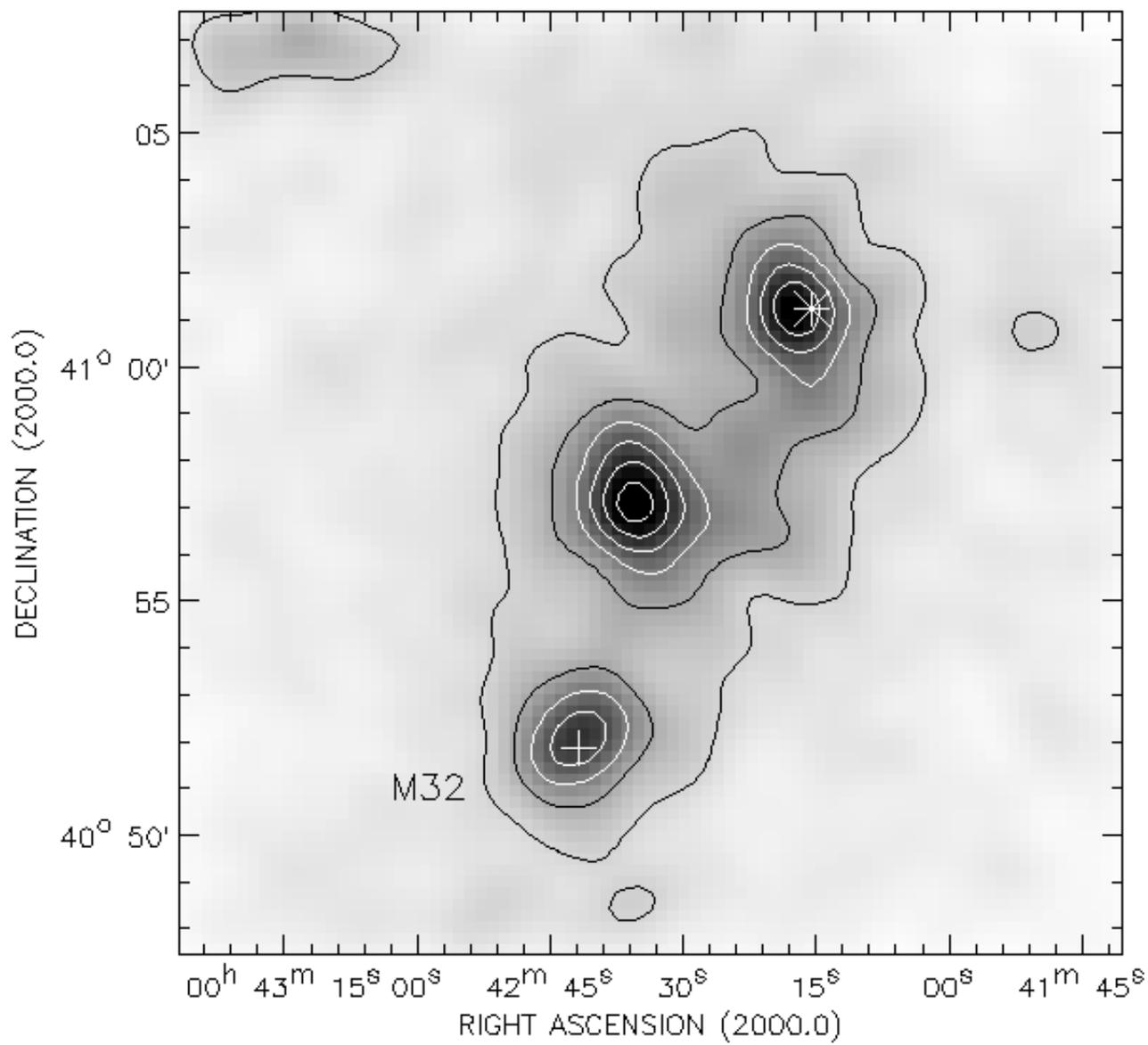
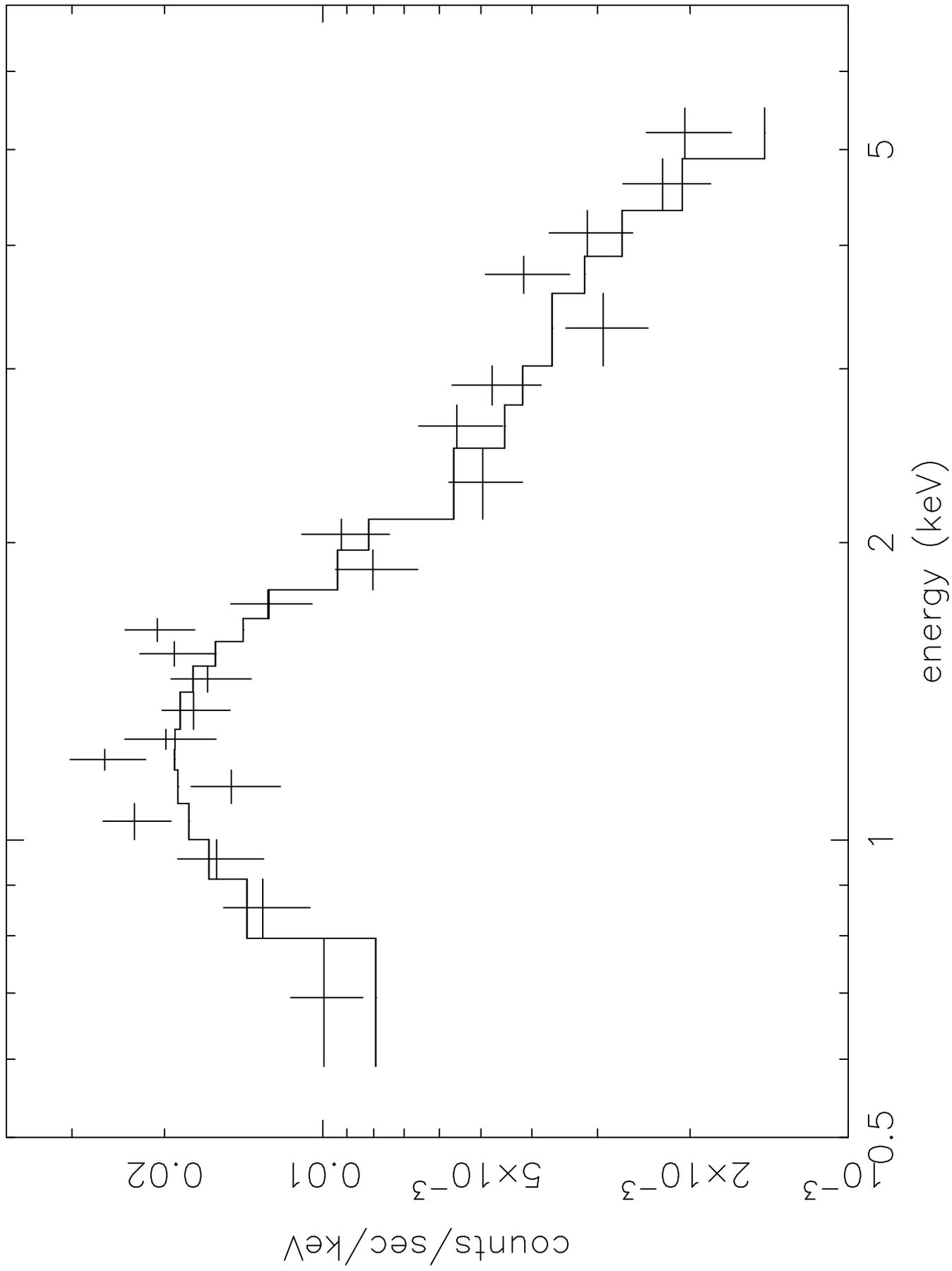
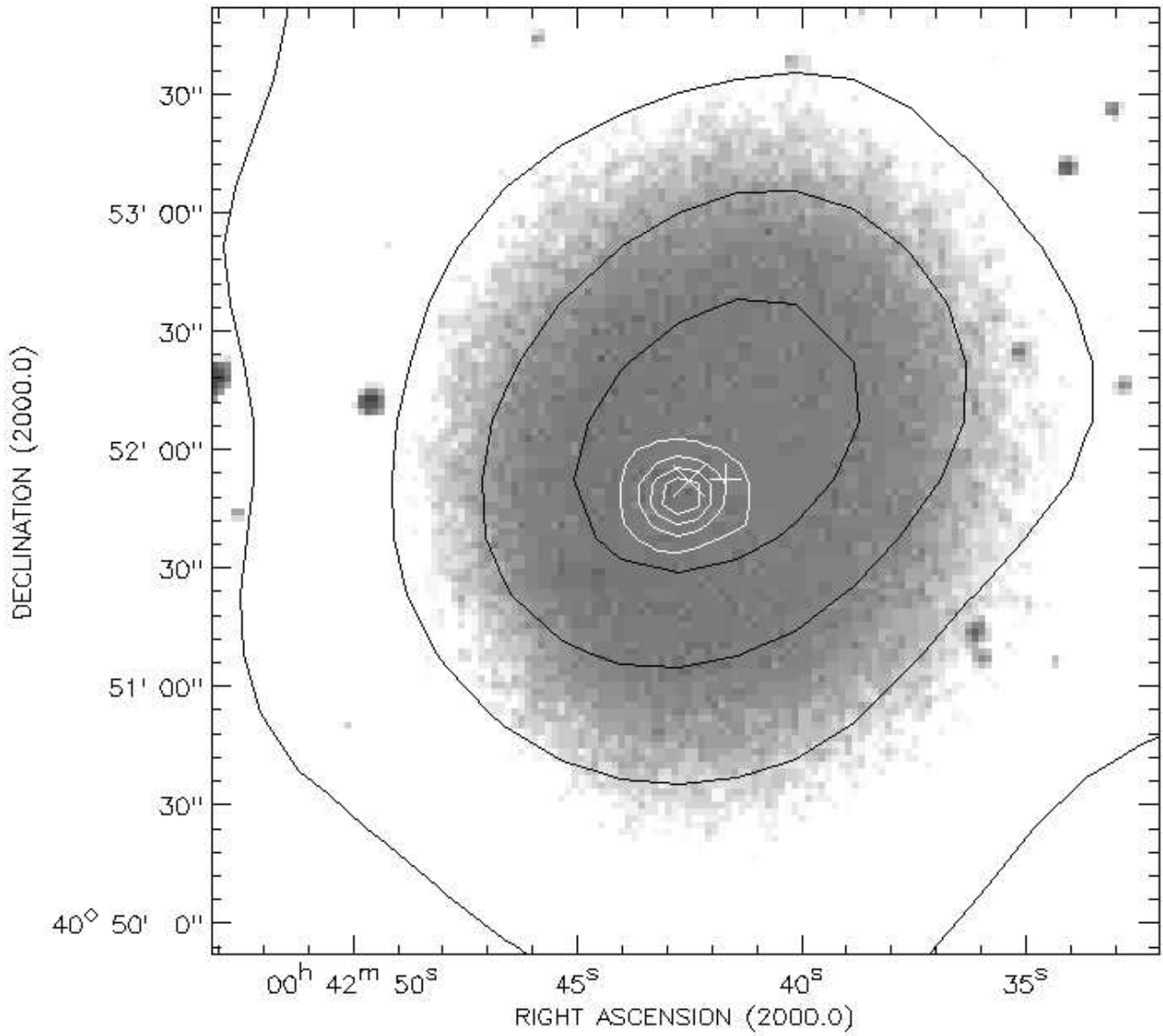
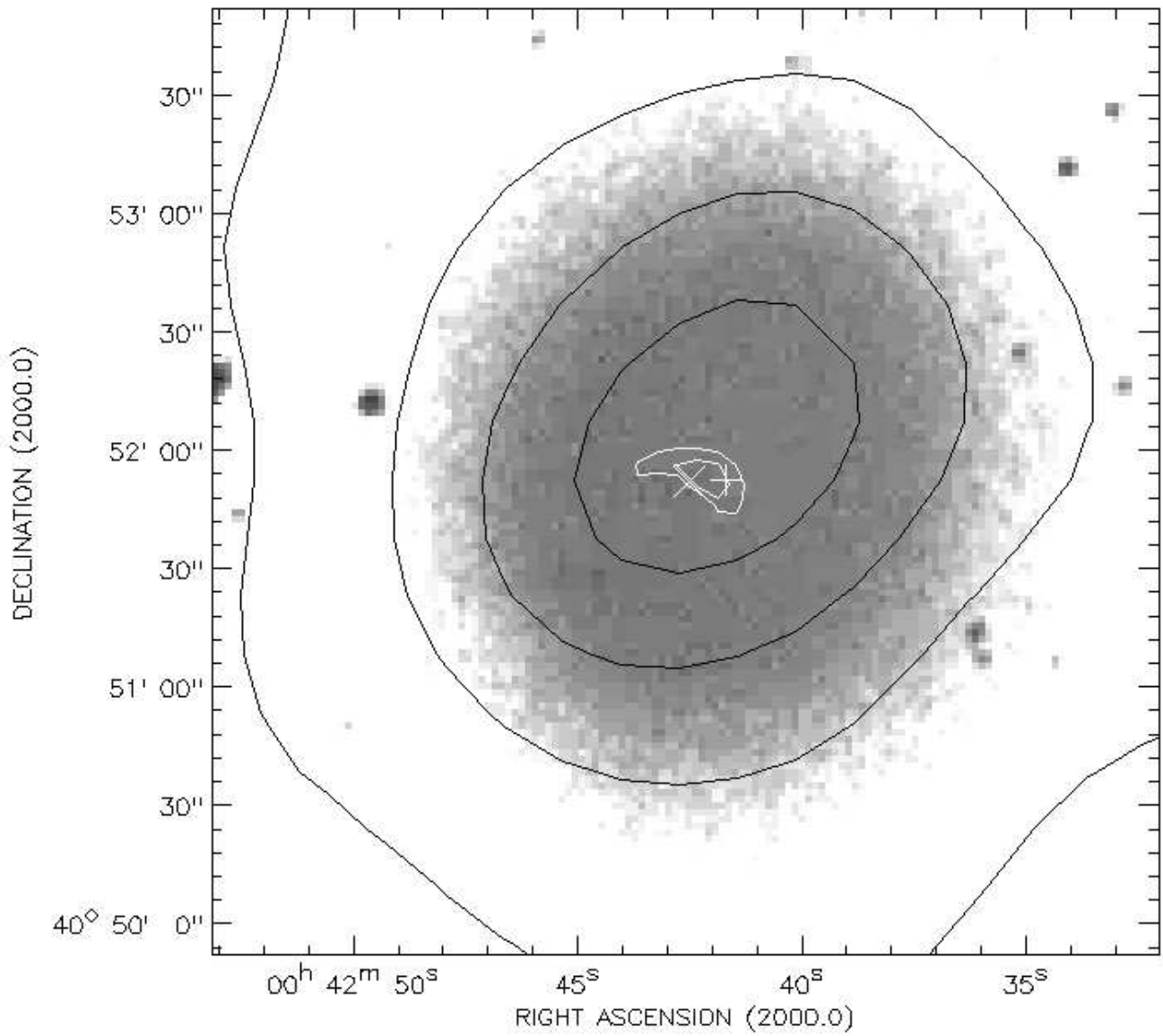
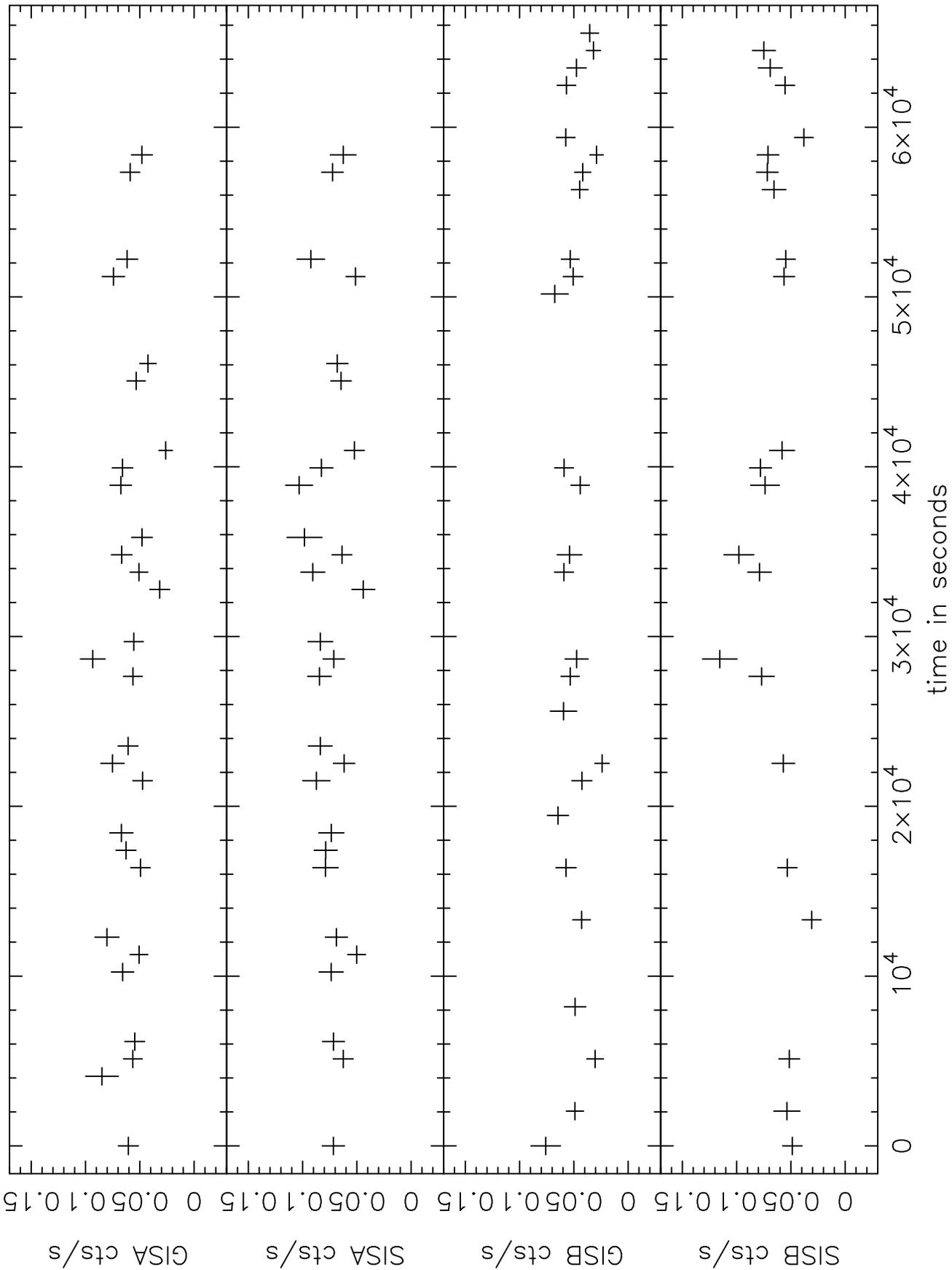
\end{document}